\documentclass[journal]{IEEEtran}

\usepackage[usenames, dvipsnames]{color}
\usepackage{multirow}
\usepackage{booktabs}
\usepackage{amsmath}
\usepackage{amsfonts}
\usepackage{multirow}
\usepackage{epsfig}
\usepackage{psfrag}
\usepackage{subfigure}
\usepackage{pifont}
\usepackage{cite}
\usepackage{url}
\newcounter{numberlistc}
\usepackage{algorithm}
\usepackage{algorithmic}
\usepackage{graphicx}
\graphicspath{ {./figs/} }

\usepackage[usenames]{color}
\usepackage{amssymb,amsmath}
\usepackage{amsthm}
\usepackage{subfigure}

\newcounter{itemlistc}

\newenvironment{itemlist}
    {   \setcounter{itemlistc}{0}
    \begin{list}{$\bullet$}
        {\usecounter{itemlistc}
        \setlength{\parsep}{0pt}
        \setlength{\topsep}{3pt}
        \setlength{\itemsep}{0pt}}
        }{ \end{list} }

\usepackage{setspace}

\usepackage[font=small, labelfont=bf]{caption}

\begin{document}

\title{GPU-based Ising Computing for Solving Balanced Min-Cut Graph
  Partitioning Problem}

\date{}

\author{
  \indent Chase Cook~\IEEEmembership{Student Member,~IEEE}, Wentian
  Jin, ~\IEEEmembership{Student Member,~IEEE}, 
  Sheldon
  X.-D. Tan~\IEEEmembership{Senior Member,~IEEE}
~\thanks{
\indent Chase Cook, Wenjin Jin and Sheldon X.-D Tan are with the Department of
Electrical and Computer Engineering,  University of California,
Riverside, CA 92521.
}
}






\maketitle

\begin{abstract}

  Ising computing provides a new computing paradigm for many hard
  combinatorial optimization problems. Ising computing essentially
  tries to solve the quadratic unconstrained binary optimization
  problem, which is also described by the Ising spin glass model and
  is also the basis for so-called Quantum Annealing computers. In this
  work, we propose a novel General Purpose Graphics Processing Unit
  (GPGPU) solver for the balanced min-cut graph partitioning problem,
  which has many applications in the area of design automation and
  others. Ising model solvers for the balanced min-cut partitioning
  problem have been proposed in the past. However, they have rarely
  been demonstrated in existing quantum computers for many meaningful
  problem sizes. One difficulty is the fact that the balancing
  constraint in the balanced min-cut problem can result in a complete
  graph in the Ising model, which makes each local update a global
  update. Such global update from each GPU thread will diminish the
  efficiency of GPU computing, which favors many localized memory
  accesses for each thread. To mitigate this problem, we propose an
  novel Global Decoupled Ising (GDI) model and the corresponding
  annealing algorithm, in which the local update is still preserved to
  maintain the efficiency. As a result, the new Ising solver
  essentially eliminates the need for the fully connected graph and
  will use a more efficient method to track and update global balance
  without sacrificing cut quality. Experimental results show that the
  proposed Ising-based min-cut partitioning method outperforms the
  state of art partitioning tool, METIS, on G-set graph benchmarks in
  terms of partitioning quality with similar CPU/GPU times.

 \end{abstract}

 \section{Introduction}
\label{sec:intro}

  Quantum computing promises vast computational power through its
  massive scalability and has prompted an increasingly large landscape
  of emerging research areas. One of these proposed areas is that of
  Quantum Annealing
  computers~\cite{MJohnson:Nature'11,BSergio:NatPhys'14} and has been
  realized in the D-Wave quantum computer~\cite{Dwave}. This paradigm
  of computing uses quantum interactions to find minimal energy states
  of a model which corresponds to the solutions of a problem. One such
  model is the Ising spin glass model. This model is a statistical
  model that describes the ferromagnetic interactions of so-called
  spin glasses and is explained in detail in
  sec.~\ref{sec:ising_algorithm}. The ground state of the model corresponds
  to optimal solutions of quadratic unconstrained binary optimization
  problems, which many hard computational problems can be mapped
  to~\cite{MJohnson:Nature'11,ALucas:FIP'14}. However, quantum
  computers and quantum annealers have yet to reach maturity and the
  costs associated with developing and deploying one of these machines
  is extremely high. As a result, adapting traditional computing
  platforms, specifically those with highly parallel compute
  capabilities, to solve the Ising model has gained traction as a
  method of implementing and showing the practicality of quantum
  annealing applications.

  The application areas for Quantum Annealing (QA) and the Ising model
  are numerous and include many hard combinatorial optimization
  problems such as max-flow, max-cut, graph partitioning,
  satisfiability, and tree based problems, which are important in many
  scientific and engineering areas. One particular problem area is
  VLSI physical design, where QA can help find optimal solutions for
  cell placement, wire routing, logic minimization, via minimization,
  and many others.  The vast complexity of modern integrated circuits
  (ICs), some having millions or even billions of integrated devices,
  means that optimal solutions to these problems are almost always
  computationally intractable and require heuristic and analytical
  methods to find approximate solutions.  It is well-known that
  traditional von Neumann based computing can not deterministically
  find polynomial time solutions to these hard
  problems~\cite{Papadimitriou:Book'98}, which shows the importance of
  non-traditional compute methods such as QA.

It has been shown that many hard combinatorial optimization problems
can be mapped to the Ising model~\cite{ALucas:FIP'14}. However, there
are many challenges to handle practical problems using the Ising
model. For instance, for the balanced min-cut partitioning problem,
the balance constraint will lead to a complete graph in the resulting
Ising model. The reason is that the {\it balance} constraint is a
global constraint. As a result, each Ising spin glass is connected to
all the Ising spins glasses in the graph, therefore; each local spin
update becomes a global update. Current Ising model solvers and
quantum annealing computers often do not have architectures amenable
to the embedding of problems with complete graphs, e.g., the Chimera
graph architecture used in D-wave computer~\cite{Dwave}.  However,
recent study shows that balanced min-cut problems on the D-wave
computer indeed yields better results than the state of the art
partitioning solvers like METIS~\cite{Mwesigwa:PMES'17}, but the
problem sizes solved are still limited to thousands of nodes and
requires co-processing on traditional hardware due to the limited
number of available qubits in the QA machine.

In this work, we apply Ising computing to solve a more practical
combinatorial problem -- the balanced min-cut graph partitioning
problem. We will first outline the Ising model and how a practical
problem like the balanced min-cut partitioning problem can be mapped
to it. We have the following contributions:

\begin{itemlist}

\item We will first present a standard implementation of the
  Ising annealing solution for 2-way balanced graph partitioning
  problem on the GPU platform. Due to the global balance constraint,
  it will lead to a complete graph in the Ising model, which requires
  each spin glass to perform a global update. Such global update from
  each spin glass will  diminish the efficiency of GPU computing, which
  favors many localized memory accesses for each thread.

\item To mitigate complete graph problem, we propose an novel
  balance-constraint efficient globally decoupled Ising (GDI) model
  and the corresponding annealing algorithm, in which the local update
  is still preserved to maintain the efficiency. As a result, the new
  Ising solver essentially eliminates the need for the fully connected
  graph and will use a more efficient method to track and update
  global balance without sacrificing cut quality.  The proposed
  methods will utilize an asynchronous update schemes to ensure
  uncorrelated spin updates and, in conjunction with decaying random
  flips, will naturally add energy to the model that will help to
  escape local minima in the solution.

\item We show that the proposed 2-way min-cut graph partitioning Ising
  solver produces cut results that compete favorably with the state of
  the art graph partitioning software METIS~\cite{METIS:JSC'99}, a
  heuristic with close to linear time complexity. Our numerical
  results on the published G-set benchmarks further show that the
  proposed Ising solver method will vastly decrease computation time
  compared to the standard method, will achieve better cut results
  (lower cut values and perfectly balanced partitions) than METIS on
  many large graph problems, and will find these solution in a
  comparable time to METIS.

\end{itemlist}

This article is organized as follows. Section~\ref{sec:related_work}
reviews some related works for quantum computing and other recently
proposed hardware based Ising
machines. Section~\ref{sec:ising_algorithm} reviews the basic concepts
of Ising models and the annealing method for Ising
models. Section~\ref{sec:partitioning} proposes two Ising solvers for
the balanced min-cut graph partitioning problem.  The graph
partitioning results and comparison against METIS and discussions are
presented in Section~\ref{sec:results}. At last,
Section~\ref{sec:conclusion} concludes this article.

 \section{Background and related work}
\label{sec:related_work}
 
While quantum computing has yet to reach maturity, there exists a
number of other hardware-based Ising model solvers which have been
proposed to exploit the highly parallel nature of this model. On the
other hand, the adiabatic process is not just limited to the quantum
devices and systems. Such natural adiabatic computing can be found in
many devices and systems. The so-called {\it natural computing} maps
the problems to natural phenomena characterized by intrinsic
convergence properties. In~\cite{MYamaoka:IJSSC'16}, a novel CMOS
based annealing solver was proposed in which an SRAM cell is used to
represent each spin and thermal annealing process was emulated to find
the ground state. In \cite{HGyoten:IEICE'18,CYoshimura:IJNC'17}, the
FPGA-based Ising computing solver has been proposed to implement the
simulated annealing process. However, these hardware Ising based
solvers suffer from several problems. First, the Ising model for many
practical problems can lead to very large connections among Ising
spins or cells. Furthermore, embedding those connections into the
2-dimensional fixed degree spin arrays in VLSI chips is not a trivial
problem and requires mitigation techniques such as cell cloning and
splitting as proposed in~\cite{HGyoten:IEICE'18,CYoshimura:IJNC'17}.
Second, ASIC implementations are not flexible and can only handle a
specific problem and FPGA implementations require architectural
redesign for different problems. Third, one has to design hardware for
the random number generator for each spin cell and simulate the
temperature changes, which has significant chip area costs which
resulting in scalability degradation. An optical parametric oscillator
based Ising machine was recently
proposed~\cite{McMahon'IEEESpectrum'18} in which the annealing is
carried out by interaction of laser pulses in the optical fibers. Also
electronic Oscillator-based Ising machine implemented by using CMOS
electronic circuits was demonstrated on max-cut
problems~\cite{WangRoychowdhury:DAC'19}. Again this work can only
solve very small and simple problems as it uses discrete electronic
components.

Based on the above observations and the highly parallel nature of the
Ising model, in this work, we propose using the General Purpose
Graphics Processing Unit (GPGPU or, more simply GPU) as the Ising
model annealing computing platform. The GPU is a general computing
platform, which can provide much more flexibility over VLSI hardware
based annealing solutions as a GPU can be programmed in a more general
way, enabling it to handle any problem that can be mapped to the Ising
model. The GPU is an architecture that utilizes large amounts of
compute cores to achieve high throughput performance. This allows for
very good performance when computing algorithms that are amenable to
parallel computation while also having very large data sets which can
occupy the computational resources of the
GPU~\cite{cuda,NvidiaKepler}. Previous Ising model solvers on GPUs
have been proposed
already~\cite{BBlock:CPC'10,LBarash:CPC'17,MWeigel:JCP'12}. However,
they still focus on small physics problems which assume a nearest
neighbor model only. Their models, are amenable to the GPU computing
as it is easily load balanced across threads but is not general enough
to handle practical combinatorial optimization problems in EDA and
other fields. Furthermore, many GPU-based methods use a checkerboard
update scheme, but this is still only practical for the nearest
neighbor model without using complicated graph embedding.

Recently the GPU-based Ising computing solution was proposed for
solving the max-cut combinatorial
problem~\cite{CookZhao:Integration'19}. The resulting Ising solver
show many orders of magnitude speedup over IBM CPLEX mathematical
programming solver~\cite{ibm_cplex} while with even better cut
results. At the same time, the Ising max-cut solver has $2000\times$
speed-up over CPU-based Ising solver for many large cases.  However,
the max-cut problem has a simple mapping to the Ising model and more
difficult problems with more complicated constraints are yet to be
explored to show that the GPU-based Ising computing can be used for
many practical combinatorial optimization problems.

\section{Ising model and Ising computing}
\label{sec:ising_algorithm}
     
\subsection{Ising model overview}
\label{sec:ising}

\begin{figure}[!ht]
  \centering
  \includegraphics[width=.55\columnwidth]{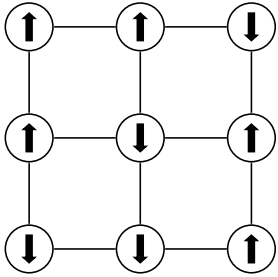}  
  \caption{The 2D nearest neighbor Ising model.}
  \label{fig:ising_model}
\end{figure}

The Ising model is a mathematical model of the ferromagnetic
interactions between so-called ``spin glasses'' or just ``spins''
inside of a two dimensional lattice. In this model, each spin is
connected to neighbors by weighted edges and the model itself is
affected by some external magnetic force or bias and is depicted in
Fig.~\ref{fig:ising_model}. Finding the minimal energy state of this
model, known as the ground state, is an NP-hard
problem~\cite{FBarahona:JPMATH'82}. However, the nature of this model
makes finding the ground state through heuristic methods highly
amenable to fine grained parallelism. Thus it is highly desirable to
map other computationally intractable problems to the problem of
finding minimal energy states of the Ising model.
 
The value of each spin $\sigma_i$ can be discrete values of $1$ or
$-1$, i.e.,  $\sigma_i \in \{-1,1\}$. The energy of the model, or the
Hamiltonian, can be defined by~\eqref{eq:ising_model_local}:

\begin{equation}
  \mathcal{H}_{i}(\sigma_{i}) =- \sum\limits_{j}^{} J_{i,j} \sigma_{i} \sigma_{j} - h_i \sigma_{i}
 \label{eq:ising_model_local}
\end{equation}

In the preceding equation $J_{i,j}$ is the interaction weight between
a spin $\sigma_i$ and its neighbor $\sigma_j$ while $h_i$ describes
the external force acting on the spin glass.

To determine the spin value of a glass $\sigma_i$, we can find its
local energy $\mathcal{H}_(sigma_i)$. To do this, we rewrite \eqref{eq:ising_model_local} as:

\begin{equation}
    \mathcal{H}_{i}(\sigma_i) = \left( -\sum\limits_{j}^{} J_{i,j} \sigma_{j} -
h_i \right) \sigma_i = - S \times \sigma_i
 \label{eq:ising_model_local_1}
\end{equation}

We can see from~\eqref{eq:ising_model_local_1} that the spin value of
a glass can actually be determined from the sign of $S$, e.g., if $S >
0$ then $\sigma_i = 1$, else if $S < 0$ then $\sigma_i = -1$ and if $S
= 0$ we allow the spin value to be random. This process describes the
{\bf local spin update}.  We note that every spin glass only requires
knowledge of connected neighbors to minimize its own local energy. The
only restriction on this is that the updates for each individual spin
must not be correlated. If this is satisfied, then each spin can be
updated independently and in
parallel~\cite{MYamaoka:IJSSC'16,HGyoten:IEICE'18,CYoshimura:IJNC'17,BBlock:CPC'10,LBarash:CPC'17}.
Then the global energy of the whole Ising model is given by the
following \eqref{eq:ising_model_global},

\begin{equation}
    \mathcal{H} (\sigma_1, \sigma_2, ..., \sigma_n) = -\sum\limits_{\langle i,j\rangle}^{} J_{i,j} \sigma_{i} \sigma_{j} - \sum\limits_{i} h_i \sigma_i
\label{eq:ising_model_global}
\end{equation}

In this equation, $\langle i,j \rangle$ refers to every combination
of spin glass interactions. This means that minimizing the local
energy of each spin glass will lead to the minimum global energy as
well. Finding this ground state is equivalent to
solving an unconstrained Boolean optimization
problem\cite{Boros:2007:LSH:1231244.1231283}. This is useful since it
has been shown that many intractable problems can be mapped to this
model~\cite{ALucas:FIP'14}.

\begin{figure}[!ht]
  \centering
  \includegraphics[width=0.57\columnwidth]{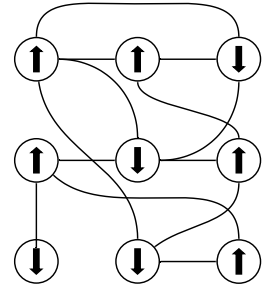}
  \caption{An example of a generally connected Ising model.}
  \label{fig:ising_model_gen}
\end{figure}

Previous implementation methods that solve the Ising model rely on the
use of the nearest neighbor model as shown in
Fig.~\ref{fig:ising_model}. However, this requires the use of graph
embedding, which is an NP-hard problem, to successfully map practical
problems to the Ising model~\cite{HGyoten:IEICE'18}. This practice is
required for other solution methods using ASIC and FPGA
implementations which are less flexible. In contrast, our use of the
GPGPU allows us to directly use a generally connected Ising model, as
shown in Fig.~\ref{fig:ising_model_gen}, to easily accommodate complex
problem cases.


\subsection{Annealing method for Ising model solution}
\label{sec:algorithm}

\begin{figure}[ht!]
  \centering
  \includegraphics[width=0.8\columnwidth]{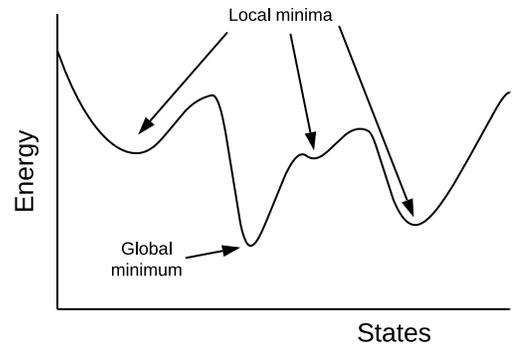}
  \caption{Depiction of the local minima and global minimum in the energy minimization problem.}
  \label{fig:energy}
\end{figure}

A trivial solution approach to solving the Ising model would be to
utilize the classical thermal annealing approach, Simulated Annealing
(SA)~\cite{Kirkpatrick:Science'83}. This method simulated the thermal
annealing process by creating a high ``temperature'' environment that
allows the heuristic to escape local minima during the search for the
minimum global energy state as depicted in Fig.~\ref{fig:energy}. As
the heuristic runs, the temperature is gradually decreased allowing it
to settle on a final solution state. While this may be a viable
method to solve the Ising model, it requires sequential updates and global energy calculations
at every step, which is costly, and unnecessary for the standard Ising model.

To find the ground state of the Ising model in this work, we utilized
a modified annealing algorithm that better mimics a QA computer and
also avoids the global energy calculation as shown in
Algorithm~\ref{alg:annealing}. This algorithm has the added benefit of
allowing us to exploit the parallelism in the local spin glass
updates, which we will introduce in a later section, while also
allowing us to avoid local minima. The algorithm allows each spin
glass to update by independently minimizing its own local energy
without considering global energy. Furthermore, to avoid the local
minima, spin glasses are randomly flipped with a gradually decaying
probability, e.g., each spin has a $50\%$ probability to flip at the
start of simulation, this probability will exponentially decrease
during simulation.

\begin{algorithm}
  \caption{Modified Ising annealing algorithm}
  \label{alg:annealing}
  \begin{algorithmic}[1]
    \STATE {input: ($M$, $N$, ${\bf S}$)}
    \STATE{initialize all $\sigma_i$ in ${\bf S}$}
    \FOR{sweep-id \textbf{in} \{1, 2, \ldots, $M$\}}
      \FOR{$\sigma_i$ \textbf{in} ${\bf S}$}
      \STATE{$\sigma_i \leftarrow \mathrm{argmin}(H(\sigma_i))$  based on  \eqref{eq:ising_model_local_1}}
      \ENDFOR
      \STATE{randomly choose and flip $N$ spin glasses in ${\bf S}$}
      \STATE{decrease $N$}
    \ENDFOR  
  \end{algorithmic}
\end{algorithm}

In this algorithm we have three parameters: $M$ which is the number of
update ``sweeps'', $N$ which controls the spin glasses to randomly
flip, and $S$ which is the set of all spin glasses $\sigma_i$ in our
model. We start by initializing all of our spin glasses $\sigma_i$ in
our model(graph) $S$. Then, $M$ sweeps are calculated. During each
sweep, every spin glass will chose its spin value
$\sigma_i \in \{1,-1\}$ such that its local energy is minimized
according to its interactions with its neighbors. At the end of an
update sweep, $N$ spin glasses are randomly flipped and $N$ decays
according to a desired decay schedule such as an exponential
decay. This process is repeated $M$ times. During this process, as
each spin glass minimizes its local energy, the global energy will
also decrease. The random flips are introduced at each phase to avoid
local energy minima and are decayed to allow the states to settle at
the end of the simulation.

We want emphasize that this Ising annealing process is fundamentally
different from the traditional simulated annealing
process~\cite{Kirkpatrick:Science'83}, in which we have to measure or
compute the global cost function for each move and make the
statistical decision (accept or rejection), which is affected by the
temperature. For Ising annealing, all the moves are local as shown in
step 5 in {\bf Algorithm 1}. As a result, steps 3 to 7 can be done in
parallel for all the cells, which indicates the massive fine-grained
parallelism, which in general is not possible in traditional
simulated annealing process. The only restriction on this parallelism is that
spin updates cannot be correlated~\cite{DLandau:book'05}.

\section{Ising model for balanced min-cut partitioning}
\label{sec:partitioning}
   

In previous works, the Ising model has been used to solve the max-cut
problem~\cite{MYamaoka:IJSSC'16,HGyoten:IEICE'18,CYoshimura:IJNC'17,CookZhao:Integration'19},
but as pointed out in~\cite{ALucas:FIP'14}, many other NP-complete and
NP-hard problems can be mapped to this model as well. The choice to
use max-cut in these other works stemmed from the fact that it is
trivial to map to the Ising model, making it ideal as a proof of
concept problem. However, very little work has been done to map and
solve other, more practical problems.

For this reason, this article will show that the balance min-cut graph
partitioning problem (min-cut) can be successfully mapped to the Ising
model and solved with solution quality rivaling that of state
of the art partitioners such as the METIS solver~\cite{METIS:JSC'99}. This
problem is highly practical as partitioning is a key algorithm in a
wide variety of applications such as VLSI Physical Design.

\subsection{Min-cut Hamiltonian}

To formulate the Hamiltonian for the min-cut problem, we must first
define min-cut. Given a graph $G(V,E)$, with edge set $E$ and
vertex set $V$, partition $G$ into two subsets ($V_1,V_2$) such that
the edges between $V_1$ and $V_2$ are minimized and that $|V_1| =
|V_2|$. In other words, we want to assign each vertex in the graph
to one of two sets such that we minimize the connections between the two sets
while also keeping the size of each set equal. Unfortunately, this
problem has been shown to be NP-hard~\cite{GareyJohnson:book'1979}.

To facilitate the formulation of the Hamiltonian and subsequent
mapping to the Ising model, we need to modify certain constraints so
that we can put the problem in a form that is amenable to
unconstrained binary optimization. To do this, we will define our
Hamiltonian to be the summation of two separate Hamiltonian functions,
$\mathcal{H}_A$ and $\mathcal{H}_B$. Minimizing the energy of this
Hamiltonian will be equivalent to solving the min-cut problem.
Furthermore, we will relax the constraint that $|V_1| = |V_2|$ and
allow some minor imbalance between each set (which we will now call
partitions). Lastly, we will define the spins in our Ising model to be
the indication of which partition a spin glass belongs in, i.e., if
$\sigma_i = 1$ then $\sigma_i \in V_1$ and if $\sigma_i = -1$, then
$sigma_i \in V_2$.
\begin{equation}
  \centering
    \mathcal{H}_{cut} = \mathcal{H}_{cut,A} + \mathcal{H}_{cut,B}
    \label{eq:mincut_hamil}
\end{equation}

The first Hamiltonian $\mathcal{H}_{cut,A}$ is a penalty function that
adds energy to our system in~\eqref{eq:mincut_hamil} when the
partition sizes are not equal and is defined as:

\begin{equation}
\centering
\mathcal{H}_{cut,A} = A((\sum\limits_{i}^{} \sigma_{i})^{2})
\label{eq:balancing}
\end{equation}

Recall that $\sigma_i \in {-1,1}$, then we can see from
\eqref{eq:balancing} that it will equal zero (be minimized) when there
are an equal number of negative and positive spins. Otherwise, this
Hamiltonian will evaluate to a positive number and add energy to the
system. The constant $A$ is a parameter that can be tuned to affect
the weight of the penalty incurred from imbalance.

The Hamiltonian $\mathcal{H}_{cut,B}$ is responsible for partitioning
the spin glasses such that the edges between the partitions are
minimized and is defined as:
\begin{equation}
  \centering
  \mathcal{H}_{cut,B} = B(\sum\limits_{\langle i,j\rangle}^{}
  \frac{1-\sigma_i\sigma_j}{2})
\label{eq:cutting}
\end{equation}

In \eqref{eq:cutting} we can see that when $\sigma_i = \sigma_j$ the
energy is minimal and the effect is opposite when this is not the
case. Effectively, this means that each vertex will produce the least
amount of energy when it is placed in the same partition that the
majority of its neighbors occupy. The parameter $B$ can be tuned to
increase or decrease the penalty of this Hamiltonian.

In general, when choosing the value of $A$ and $B$, the following
rule, as outlined in~\cite{ALucas:FIP'14}, can be applied:
\begin{equation}
  \centering
  \frac{A}{B} \geq \frac{min(2\Delta,N)}{8}
  \label{eq:constnat_criteria}
\end{equation}
 
In this equation, $\Delta$ is the maximum degree of the graph $G$ and
$N$ is the number of spins used to encode the problem, which for our
implementation is just the number of vertices in $G$.

By minimizing the energy of~\eqref{eq:mincut_hamil}, we can obtain a
balanced partitioning of a graph problem while also achieving
minimized cuts between the partitions. Unlike the traditional Ising
model, this Hamiltonian contains two sub-Hamiltonians. Furthermore, we
also must note that the Hamiltonian in~\eqref{eq:balancing} requires
that each spin-glass have knowledge of every spin glass in the
model. This effectively produces a fully connected graph and vastly
increases the complexity of the problem, which will be addressed in the
following sections.

\section{GPU-based Ising solver for balanced min-cut problem}
\label{sec:gpu_impl}
  

\subsection{GPU architecture}
The general purpose GPU is an architecture designed for highly
parallel workloads which is leveraged by Nvidia's CUDA, Compute
Unified Device Architecture, programming model~\cite{cuda}.  The
Nvidia GPU architecture is comprised of several Symmetric
Multiprocessors (SMs), each containing a number of ``CUDA'' cores, and
a very large amount of DRAM global memory~\cite{NvidiaKepler}. The
Kepler architecture based Tesla K40c GPU, for example, has 15 SMs for
a total of 2880 CUDA cores (192 cores per SM), and 12GB DRAM global
memory. Additionally, each SM has additional special function units,
shared memory, and cache. Compared to a multi-core CPU, the GPU offers
much more parallel computational resources at the expense of higher
latency. That is, it can do more at the same time but each operation
takes longer compared to the CPU. In general, as long as the data set
can occupy the large resources of the GPU and each operation being
performed in parallel is not latency constrained, the GPU will see
significant performance gains over the CPU.

\begin{figure}
  \centering
  \resizebox{.4\textwidth}{!}{\input{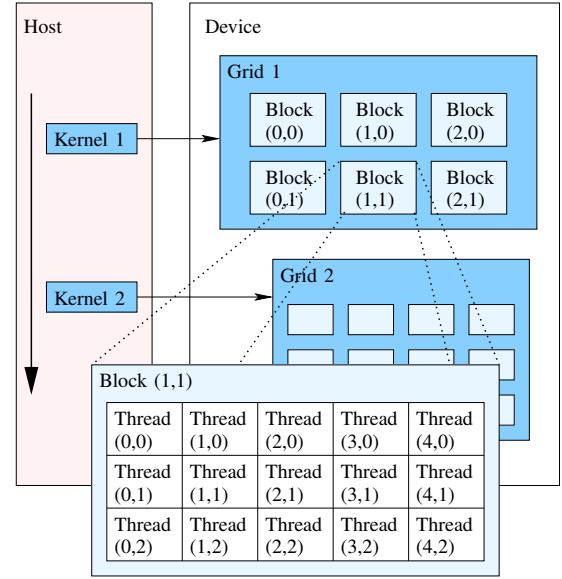}}
  \caption{The Nvidia CUDA programming model}
  \label{fig:cuda_model}
\end{figure}

The CUDA programming model, shown in Fig.~\ref{fig:cuda_model},
extends the C language adding support for thread and memory allocation
and also the essential functions for driving the
GPU~\cite{CUDA_C_Programming_Guide_2018}. The model makes a
distinction between the host and device or the CPU and GPU
respectively. The model uses an offloading methodology in which the
host can launch a device kernel (the actual GPU program) and also
prepare the device for the coming computation, e.g.,the host will
create the thread organization, allocate memory, and copy data to the
device. In practice, a programmer must call many threads which will be
used to execute the GPU kernel. Thread organization is therefore
extremely important in GPU programming. Threads are organized into
blocks which are organized into grids. Each block of threads also has
its own shared memory which is accessible to all the threads in that
block. Additionally, the threads in the block can also access a global
memory on the GPU which is available to all threads across all blocks.

    

The GPU fundamentally focuses on throughput over speed. This
throughput is achieved through the massive compute resources able to
be run in parallel. Because of this, it is important to realize that
the GPU is not meant for small data sets or extremely complicated
operations that may be better suited for a powerful, low latency,
CPU. Instead, the GPU is meant to execute relatively simple
instructions on massive data in parallel that can occupy the GPU
resources for an extended period of time.



\subsection{Standard implementation}

To first validate the GPU implementation of the Ising model solver
for the min-cut problem, we construct a parallel algorithm from the
one shown in Algorithm~\ref{alg:annealing} but with the modified
Hamiltonians for the min-cut problem.

As mentioned before, spin glass updates can be performed in parallel
during a sweep. However, the restriction on this is that the updates
must be performed independently. Furthermore, the updates of spins
must happen asynchronously to avoid biphasic oscillation. Fortunately,
this can be accomplished in the GPU simply by allowing race conditions
on read operations. Effectively, this just means that when a spin
calculates its own local energy, it will use the current energy of its
neighbors which may, or may not, have already minimized their own
local energy. To facilitate this, the following algorithm for the GPU
implementation has been used:

\begin{algorithm}

  \caption{Standard implementation for GPU Ising model annealing for
    balanced min-cut problem}
  \label{alg:annealing_gpu_1}
  \begin{algorithmic}
    \STATE{ input: ($M$, $P_f, {\bf S}$)}
    \STATE{initialize ALL $\sigma_i$ in ${\bf S}$}
    \FOR{sweep-id \textbf{in} \{1, 2, \ldots, $M$}
    \FOR{all $\sigma_i \in \bf S$ \textbf{in parallel}}
      \STATE{$\sigma_i\leftarrow\mathrm{argmin}(\mathcal{H}_{cut,A}(\sigma_i)+\mathcal{H}_{cut,B}(\sigma_i))$}
      \STATE{flip $\sigma_i$ with probability $P_f$}
     \ENDFOR
     \STATE{reduce $P_f$}
     \ENDFOR
  \end{algorithmic}
\end{algorithm}

In this algorithm the spins for each glass are initialized and $M$
update sweeps are performed. During each sweep, every spin glass, in
parallel, chooses the value of $\sigma_i$ that minimizes the
Hamiltonian in~\eqref{eq:mincut_hamil}. In effect, this means each
spin will read the spin values of all of its connected neighbors (to
evaluate~\eqref{eq:cutting}) as well as the spin values of all spins
in the entire graph (to evaluate~\eqref{eq:balancing}).
After a spin glass finishes its update, it will randomly flip with a
probability determined by $P_f$. Once every spin has updated and
decided to flip or not, the flip probability $P_f$ is reduced
according to a cooling schedule. This process is then repeated.

The flip probability $P_f$ is initially determined by the number of
nodes that should be randomly flipped, which we will call
$N_{rf}$. That is, if there are $500$ nodes, and the $N_{rf}$ is
$100$, then $P_f$ will initially be calculated to be $0.20$. When
deciding to flip or not, each spin generates a uniformly distributed
random number between $0.00$ and $1.00$. If this number is higher than
$P_f$ then the spin will not flip, however, if it is equal or lower,
then it will.

Practically, each spin is handled by a single thread in the GPU. If
the number of spin glasses exceeds the number of threads available,
then threads will be assigned multiple spin glasses to update. For the
random number generation, the CUDA cuRAND library is used which allows
for efficient random number generation in parallel. We also note, as
mentioned previously, that each spin update should not be
correlated. For this reason, updates are done asynchronously. In
effect, this means that a spin glass will check the status of its
neighbor spins during an update but it will not be guaranteed that its
neighbors have not already finished their own update operation.

%

\subsection{Globally Decoupled Ising implementation}

While Algorithm~\ref{alg:annealing_gpu_1} does successfully calculate
a balanced min-cut partition, as we will show in the results section,
it is also heavily constrained by the direct computation of
$\mathcal{H}_{cut,A}$ as shown in~\eqref{eq:balancing}. This means
that for each spin glass in the model, it must traverse the entire
model to assess the status of every spin glass and determine the
balancing penalty. This is obviously a heavy computational step and
will drastically limit the performance of the algorithm as the problem
size increases. {\it This is actually is one of major challenges for
  Ising based computing as many constraints will lead to dense or
  complete Ising graphs.}

One of the major contributions of this work is to find a way to
mitigate this challenging problem.  To mitigate this problem, we have
to find a better way to deal with the balance constraints instead of
using the standard Ising model. In this way, we still keep the spin
updates local (each spin glass no longer has to traverse the entire
model), which is critical to maintaining the efficiency and
scalability for the GPU-based computing, while making the global
connections decoupled. Therefore, we propose to use a Globally
Decoupled Ising (GDI) solver.

\begin{algorithm}
  \caption{Globally Decoupled Ising (GDI) implementation for  GPU Ising model annealing for balanced min-cut problem}

  \label{alg:annealing_gpu_2}
  \begin{algorithmic}
    \STATE{ input: ($M$, $P_f, {\bf S}$)}
    \STATE{initialize ALL $\sigma_i$ in ${\bf S}$}
    \STATE{$G = \sum\limits_{i}^{} \sigma_{i}$}
    \FOR{sweep-id \textbf{in} \{1, 2, \ldots, $M$}
    \FOR{all $\sigma_i \in \bf S$ \textbf{in parallel}}
      \STATE{$\sigma_i\leftarrow\mathrm{argmin}(\mathcal{H}_{cut,B}(\sigma_i)+ A(G + \sigma_i)^2)$}
      \STATE{$G = \mathrm{atomicAdd}(G + \sigma_i)$} 
      \STATE{flip $\sigma_i$ with probability $P_f$}
     \ENDFOR
     \STATE{reduce $P_f$}
     \ENDFOR
  \end{algorithmic}
\end{algorithm}

Specifically, in Algorithm~\ref{alg:annealing_gpu_2}, we pre-compute
the global balance before starting the first update sweep and store
this value in a global variable $G$. Then, as each thread performs its
update during a sweep, it uses this value to determine the balancing
penalty it's spin glass may incur. After deciding which partition will
minimize the spin glass local energy, the thread updates the $G$ value
so that all other spins glasses have knowledge of the new balance. It
should be noted that because this is happening in parallel, the
balance that some threads see may not be the actual balance when that
thread updates the spin glass it is responsible for. Furthermore, an
atomic addition is used to update $G$ which will incur some
computational penalty (but at much less cost than the method in
Algorithm~\ref{alg:annealing_gpu_1}). This atomic operation ensures
the validity and integrity of the $G$ value at the end of each sweep
as it eliminates ``data race'' between threads updating this value.

\subsection{Further discussion and comparison}

The primary difference in the two algorithms is the modification to
the calculation of the balancing Hamiltonian. The major effect is that
we no longer require each node to traverse the entire graph at each
update step to calculate the global balance. Rather, the global
balance is pre-computed and then individually read, while modified
(through atomic operations) by one node at a time. 

With respect to implementation, we notice that the standard version
requires a higher random flip probability than the  enhanced version to
achieve good cut results. The reason for this is that the enhanced
version naturally introduces more noise, or randomness, to the update
scheme while the standard version is more constrained by the global
update. That is, it is less willing to violate the global constraint
as each node has nearly perfect knowledge of the state of the global
balance. For this reason, during implementation, we typically need a
$5X$ increase to the value of $P_f$ when using the standard
implementation.

\section{Experimental results and discussions}
\label{sec:results}
  
In this section, we present the experimental results showing the
quality of our parallel GPU-based Simulated Quantum Annealing solver
for the balanced min-cut problem. The CPU-based solution is done using
a Linux server with 2 Xeon processors, each having 8 cores (2 threads
per core) and a total of 32 threads, and 72 GB of memory. On the same
server, we also implement the GPU-based solver using the Nvidia Tesla
K40c GPU which has 2880 CUDA cores and 12 GB of memory. Test problems
from the G-set benchmark~\cite{gset} are used for testing.
 

\begin{table*}[t]

  \caption{Summary of results for the standard (std.) and Globally decoupled 
      (GDI) Ising solver compared with the METIS results. Graphs with
    less than 1000 nodes are omitted}
  \centering
\resizebox{\textwidth}{!}{%
\begin{tabular}{ | l | l | l | l | l | l | l | l | l | l | l | l | l | }
\hline
	Graph-ID & \# nodes & \# edges & Density &$t_\text{Metis}$ & $t_\text{Ising,GDI}$  & $t_\text{Ising,std.}$ & $cut_\text{Metis}$ & $cut_\text{Ising,GDI}$ & $cut_\text{Ising,std}$ & $bal_\text{Metis}$ & $bal_\text{Ising,GDI}$ &$bal_\text{Ising,std.}$ \\ \hline \hline
	G70 & 10000 & 9999 & 2.0E-4        & 1.6E-2      & 0.34929 & 31.88844 & 487 & 507 & 555 & 5 & 0 & 0 \\ 
	G81 & 20000 & 40000 & 2.0001E-4    & 1.6E-2     & 0.15046 & 22.40281 & 210 & 240 & 242 & 5 & 0 & 0 \\ 
	G77 & 14000 & 28000 & 2.8573E-4    & 1.2E-2     & 0.15686 & 15.67784 & 220 & 212 & 232 & 3 & 0 & 0 \\  
	G67 & 10000 & 20000 & 4.0004E-4    & 0.08      & 0.15579 & 11.20150 & 216 & 216 & 244 & 1 & 0 & 0 \\ 
	G72 & 10000 & 20000 & 4.0004E-4    & 0.08      & 0.15707 & 11.23208 & 216 & 212 & 230 & 1 & 0 & 0 \\ 
	G66 & 9000 & 18000 & 4.4449E-4     & 8.0E-3     & 0.15810 & 10.08537 & 196 & 220 & 242 & 2 & 0 & 0 \\ 
	G65 & 8000 & 16000 & 5.0006E-4     & 8.0E-3     & 0.15766 & 9.01416 & 178 & 168 & 178 & 3 & 0 & 0 \\ 
	G62 & 7000 & 14000 & 5.7151E-4     & 8.0E-3     & 0.15730 & 7.87871 & 144 & 146 & 148 & 1 & 0 & 0 \\ 
	G60 & 7000 & 17148 & 7.0002E-4     & 2.8E-2     & 0.21494 & 9.31080 & 3330 & 3115 & 3321 & 2 & 0 & 0 \\ 
	G57 & 5000 & 10000 & 8.0016E-4     & 4.0E-3     & 0.15679 & 5.64559 & 140 & 110 & 252 & 0 & 0 & 0 \\ 
	G55 & 5000 & 12498 & 1.0001E-3     & 0.02       & 0.21558 & 6.45212 & 2463 & 2308 & 2391 & 0 & 0 & 0 \\ 
	G48 & 3000 & 6000 & 1.33378E-3     & 4.0E-3     & 0.15601 & 3.45138 & 120 & 124 & 194 & 1 & 0 & 0 \\ 
	G49 & 3000 & 6000 & 1.33378E-3     & 4.0E-3     & 0.15632 & 3.45322 & 74 & 60 & 136 & 0 & 0 & 0 \\ 
	G50 & 3000 & 6000 & 1.33378E-3     & 0           & 0.15617 & 3.45544 & 58 & 50 & 92 & 1 & 0 & 0 \\ 
	G64 & 7000 & 41459 & 1.6924E-3     & 3.2E-2      & 0.38691 & 5.89811 & 9946 & 9787 & 9967 & 2 & 2 & 0 \\ 
	G32 & 2000 & 4000 & 2.001E-3       & 0           & 0.12023 & 2.30673 & 50 & 40 & 54 & 1 & 0 & 0 \\ 
	G33 & 2000 & 4000 & 2.001E-3       & 0            & 0.12019 & 2.30543 & 54 & 50 & 78 & 1 & 0 & 0 \\ 
	G34 & 2000 & 4000 & 2.001E-3       & 0            & 0.12044 & 2.31117 & 112 & 86 & 150 & 1 & 0 & 0 \\ 
	G58 & 5000 & 29570 & 2.366E-3      & 2.4E-2        & 0.34457 & 3.98248 & 7226 & 6921 & 7302 & 1 & 0 & 0 \\ 
	G36 & 2000 & 11766 & 5.8859E-3     & 8.0E-3      & 0.23601 & 1.63955 & 2896 & 2722 & 2898 & 0 & 0 & 0 \\ 
	G35 & 2000 & 11778 & 5.8919E-3     & 1.2E-2     & 0.20426 & 1.77944 & 2942 & 2771 & 2841 & 1 & 0 & 0 \\ 
	G38 & 2000 & 11779 & 5.8924E-3     & 1.2E-2    & 0.26296 & 1.63863 & 2866 & 2688 & 2801 & 1 & 0 & 0 \\ 
	G37 & 2000 & 11785 & 5.8954E-3     & 1.2E-2    & 0.24608 & 1.63948 & 2921 & 2781 & 2834 & 1 & 0 & 0 \\ 
	G22 & 2000 & 19990 & 0.01          & 1.2E-2    & 0.15961 & 0.87029 & 6925 & 6739 & 6803 & 0 & 0 & 0 \\ 
	G23 & 2000 & 19990 & 0.01          & 1.6E-2    & 0.15989 & 0.86660 & 6946 & 6702 & 6743 & 0 & 0 & 0 \\ 
	G24 & 2000 & 19990 & 0.01          & 1.6E-2    & 0.16047 & 0.96611 & 7022 & 6719 & 6809 & 1 & 0 & 0 \\ 
	G28 & 2000 & 19990 & 0.01          & 1.2E-2    & 0.16016 & 0.86641 & 6961 & 6753 & 6781 & 0 & 0 & 0 \\ 
	G30 & 2000 & 19990 & 0.01          & 1.6E-2    & 0.16170 & 0.86767 & 6978 & 6702 & 6774 & 1 & 0 & 0 \\ 
	G31 & 2000 & 19990 & 0.01          & 1.2E-2    & 0.16086 & 1.00123 & 6957 & 6683 & 6798 & 1 & 0 & 0 \\ 
	G51 & 1000 & 5909 & 1.183E-2       & 4.0E-3    & 0.18105 & 0.92766 & 1513 & 1383 & 1458 & 0 & 0 & 0 \\ 
	G53 & 1000 & 5914 & 1.184E-2       & 4.0E-3    & 0.19386 & 0.92421 & 1498 & 1366 & 1368 & 0 & 0 & 0 \\ 
	G52 & 1000 & 5916 & 1.1844E-2      & 4.0E-3    & 0.19584 & 0.92683 & 1446 & 1394 & 1399 & 0 & 0 & 0 \\  
	G54 & 1000 & 5916 & 1.1844E-2      & 4.0E-3    & 0.20007 & 0.92612 & 1465 & 1356 & 1430 & 0 & 0 & 0 \\ 
	G43 & 1000 & 9990 & 0.02           & 8.0E-3    & 0.15092 & 0.51411 & 3542 & 3350 & 3382 & 0 & 0 & 0 \\ 
	G44 & 1000 & 9990 & 0.02           & 8.0E-3 & 0.15365 & 0.51538 & 3565 & 3361 & 3402 & 0 & 0 & 0 \\ 
	G45 & 1000 & 9990 & 0.02           & 8.0E-3 & 0.15322 & 0.51169 & 3522 & 3347 & 3397 & 0 & 0 & 0 \\ 
	G46 & 1000 & 9990 & 0.02           & 4.0E-3 & 0.15210 & 0.51384 & 3573 & 3353 & 3386 & 0 & 0 & 0 \\ 
	G47 & 1000 & 9990 & 0.02           & 8.0E-3 & 0.15191 & 0.51398 & 3520 & 3350 & 3396 & 0 & 0 & 0 \\ 
\hline
\end{tabular}
}

  \label{tb:summary}
\end{table*}

\subsection{Solution time study}

We firstly investigate the performance of the standard Ising solver and
the globally decoupled method developed in the paper, followed by discussion of
its performance compared to the state of the art partitioning software.

The direct Ising solver implementation of the min-cut partitioning
problem leads to a complete graph requiring a global update for each
spin glass in the model. In other words, when each node updates during
an annealing sweep, it must visit every other node in the graph. To
address this, we proposed the balance constraint efficient annealing
algorithm in~\ref{alg:annealing_gpu_2}. In the GDI version, we
mitigate the issue of global update by utilizing a global variable to
store the balance of the graph partition which is updated using atomic
operations by each node. This means each node only needs to perform a
single read and atomic add operation on this variable.

\begin{figure}
  \centering
  \includegraphics[width=\columnwidth]{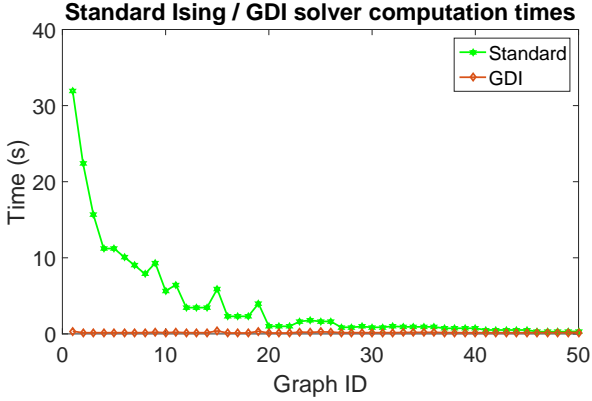}
  \caption{Performance gains of the GDI solver min-cut algorithm.}
  \label{fig:mod_sqa_v_naive_sqa}
\end{figure}

In Fig~\ref{fig:mod_sqa_v_naive_sqa}, the graph problems solved are
sorted from lowest to highest density where density is calculated as
$density = (2\times \# edges) / (\# nodes \times(\# nodes - 1))$ . The
nature of the G-set graphs used are such that low density graphs have
many more nodes and edges than the high density graphs. For this
reason, the graph can be looked at as being sorted from large
complexity to smallest complexity. As we can see, the direct
implementation quickly increases in computation time as the problem
become more and more complex. Comparatively, the GDI version
appears constant in these results as the computations times are not
comparable.

We further show the performance results of the GDI solver algorithm
and the performance of METIS, widely considered the gold standard for
partitioning software.

\begin{figure}
  \centering
  \includegraphics[width=\columnwidth]{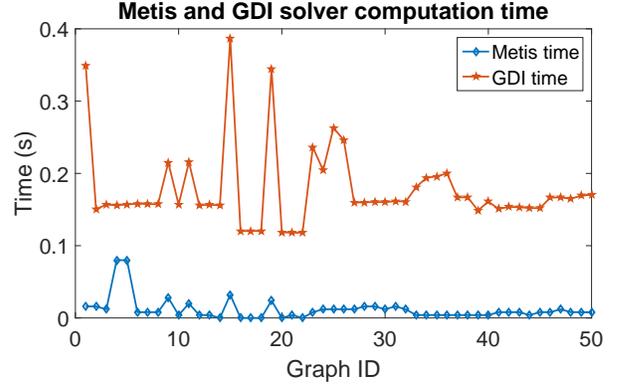}
  \caption{Performance comparison of the GDI GPU Ising solver and METIS.}
  \label{fig:sqa_v_METIS_time}
\end{figure}

We can see from Fig.~\ref{fig:sqa_v_METIS_time} that, while METIS is
still able to beat the proposed solver in terms of speed, the proposed
Ising solver is quite close and comparable in time with all solutions
taking less than a second, even for the larger graphs.

\subsection{Solution quality study}
\label{sec:quality}

To study the solution quality of the proposed method, wee first show
that the solution quality of the direct implementation of the
GPU-based Ising solver and the GDI solver are similar.

\begin{figure}
  \centering
  \includegraphics[width=\columnwidth]{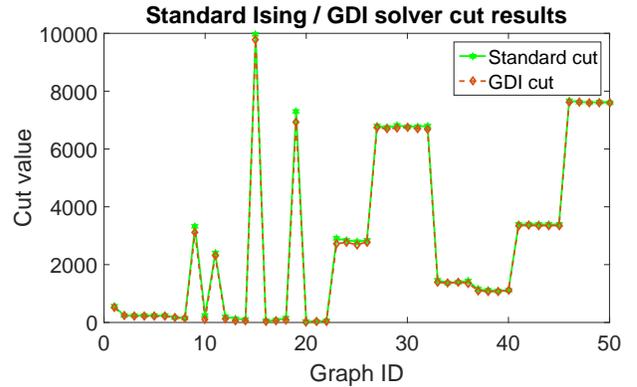}
  \caption{Solution quality of the proposed GPU-based Ising solvers.}
  \label{fig:sqa_v_sqan_quality}
\end{figure}

As seen from Fig.~\ref{fig:sqa_v_sqan_quality} the solution quality of
the GDI method not only achieves similar solution results, it
also consistently achieves cut values marginally lower than the standard
implementation.

To measure the solution quality produced by the proposed method, we
compare balanced min-cut partitioning results of the proposed method
with the results produced by METIS. We omit the solution quality
results of the direct implementation as they are similar to the
GDI method proposed. For fairest comparison, we ensured both
METIS and the GPU-based Ising solver used a highly constrained balance
criteria. For each graph, we run both solvers 10 times and use the
best solution quality found. The quality is measured by comparing the
cut values of the solvers. Cut values are defined as the number of
edges that connect one partition to the other.

\begin{figure}
  \centering
  \includegraphics[width=\columnwidth]{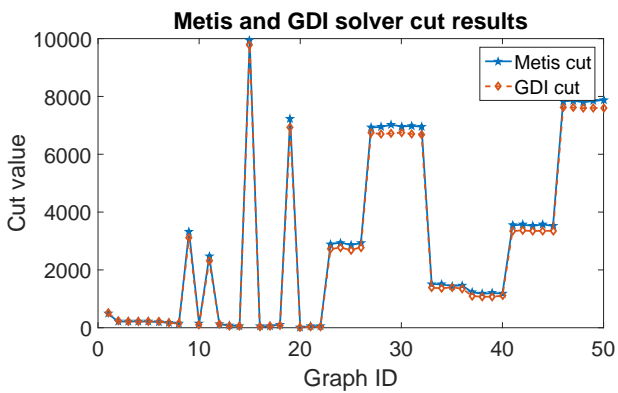}
  \caption{Cut quality results for the G-set benchmark problems.}
  \label{fig:sqa_v_METIS_quality}
\end{figure}

As seen from Fig.~\ref{fig:sqa_v_METIS_quality}, the proposed Ising solver GPU
method achieves the same or better quality results for almost every
graph tested. Of the 51 graphs, the Ising solver method achieved worse quality
in only 5 graphs and produced the exact same quality in 1 graph. Even
then, the Ising solver results were very close to METIS. We note that the all
of the graphs where Ising solver achieved worse quality were very sparse
graphs, i.e., their graph density was very low, which suggests that
utilization of the Ising model can be difficult for these types of
graphs.

While the balancing constraint on METIS was set to be as restrictive
as possible, $22$ of the the $51$ graphs tested had imbalanced
partitions. In contrast, the proposed GDI solver achieved perfect
balance for all but one graph tested. We also remark that the balances
achieved by METIS were still quite good with the worst imbalance being
only 5 nodes.

These results, as summarized in Table~\ref{tb:summary}, show that the
proposed method is able to achieve better solution quality than the
state of the art METIS solver, both in terms of balanced partitions
and minimized cut values. Furthermore, even for large and dense
graphs, the proposed solver finished in a time comparable to METIS.

\section{Conclusion}
\label{sec:concl}

In this work, we have proposed a GPU-based Ising spin glass model
solver applied to the balanced min-cut graph partitioning problem. The
work presented the Ising spin glass model and the annealing solution
method for solving Unconstrained Quadratic Binary Optimization
problems while additionally showing the method for mapping the min-cut
problem to this model. A standard GPU implementation is presented in
addition to a Globally Decoupled Ising annealing algorithm which
mitigates the costly global updates during each annealing
sweep. Numerical results show that both the standard and enhanced
annealing algorithms achieve high quality, and nearly perfectly
balanced, partitioning results that compete with and exceed the
partitioning quality achieved by the state of the art partitioning
solver, METIS. Furthermore, the proposed GDI method can produce
results much faster than the standard method, resulting in computation
times comparable to the METIS solver.  Experimental results show that
the proposed Ising-based min-cut partitioning method outperforms the
state of art partitioning tool, METIS, on G-set graph benchmark in
terms of partitioning quality with similar CPU/GPU times.

\label{sec:conclusion}


\bibliographystyle{ieeetr}
\bibliography{../../bib/physical,../../bib/security,../../bib/emergingtech,../../bib/thermal_power,../../bib/mscad_pub,../../bib/interconnect,../../bib/stochastic,../../bib/simulation,../../bib/modeling,../../bib/reduction,../../bib/misc,../../bib/architecture,../../bib/reliability,../../bib/partition}

\end{document}